\documentclass[%
aip,
amsmath,amssymb,
reprint,
]{revtex4-1}

\usepackage{graphicx}
\usepackage{dcolumn}
\usepackage{bm}

\usepackage[utf8]{inputenc}
\usepackage[T1]{fontenc}
\usepackage{mathptmx}
\usepackage{etoolbox}

\makeatletter
\def\@email#1#2{%
 \endgroup
 \patchcmd{\titleblock@produce}
  {\frontmatter@RRAPformat}
  {\frontmatter@RRAPformat{\produce@RRAP{*#1\href{mailto:#2}{#2}}}\frontmatter@RRAPformat}
  {}{}
}%
\makeatother
\begin{document}

\title{A melting mode of frozen sessile droplets with unmelted ice layer deposited at the bottom}

\author{Jiawang Cui}
\affiliation{State Key Laboratory of Engines, Tianjin University, Tianjin, 300350, China.
}%
\author{Yugang Zhao}%
\affiliation{Shanghai Key Laboratory of Multiphase Flow and Heat Transfer in Power Engineering, School of Energy and Power Engineering, University of Shanghai for Science and Technology, Shanghai 200093, China.
}%

\author{Tianyou Wang}
\affiliation{State Key Laboratory of Engines, Tianjin University, Tianjin, 300350, China.
}%
\affiliation{National Industry-Education Platform of Energy Storage, Tianjin University, Tianjin, 300350, China.
}%
\author{Zhizhao Che*}
\affiliation{State Key Laboratory of Engines, Tianjin University, Tianjin, 300350, China.
}%
\affiliation{National Industry-Education Platform of Energy Storage, Tianjin University, Tianjin, 300350, China.
}%
 \email{corresponding author: chezhizhao@tju.edu.cn}
\date{\today}

\begin{abstract}
Water-repellent properties of superhydrophobic surfaces make them promising for anti-icing and deicing applications. Through experimental visualization of frozen sessile droplets undergoing melting on superhydrophobic surfaces, we identify a melting mode with the unmelted ice layer deposited at the bottom of the melting droplet, even though the density of ice is lower than that of water. In the deposited mode of the melting process, the time required for the frozen droplet to melt completely is much shorter than that in the floating mode. Force analysis shows that the melted fluid flows along the gas-liquid interface toward the top of the melting droplet, thereby exerting force and then suppressing the upward movement of the unmelted ice layer. Moreover, the flow within the liquid film formed between the unmelted ice layer and the heating wall is dominated by the viscous force, which has a lubrication effect and maintains the deposition of the unmelted ice layer. High heating temperature, large contact angle, and low particle concentration are helpful for the occurrence of the deposited mode. 
\end{abstract}

\maketitle

Ice formation and subsequent melting in sessile droplets are key phase-change phenomena with significant implications for anti-icing and deicing. For example, when water droplets in cold air collide with an aircraft \cite{1}, the adhesion and accumulation of ice layers affect the performance and reliability of aircraft. In addition, processes such as power transmission \cite{2}, wind-power generation \cite{3}, electronic component packaging \cite{4}, phase-change energy storage \cite{5}, and 3D printing \cite{6}, all involve droplet melting phenomena. Due to the necessity of deicing, the melting process of droplets is worth exploring. In fact, many solid-liquid phase transition processes involve complex internal flow \cite{8, 7}, and the melting mode is influenced by the melting flow pattern \cite{9}. Therefore, the exploration of the melting mode needs to fully consider the situation of the melting flow under different conditions.

During the melting process of frozen droplets, the presence of an unmelted ice layer can induce a temperature gradient in the melted area. With a temperature gradient on the gas-liquid interface of the droplet, a surface-tension gradient can induce the formation of thermal Marangoni convection \cite{8, Zhao2019PIV}. In addition, with a temperature gradient on the interior of the droplet, a density gradient is generated and can induce the formation of natural convection \cite{7}. The combined effects of thermal Marangoni convection and natural convection will determine the overall flow pattern not only in the droplet freezing process \cite{Chu2019reicing, Chu2024Salty2, Chu2019Bubble} but also in the melting process of frozen droplets \cite{Chu2017melting, 9, Lei2015melting, Wang2022dewetting}. Given the practical need for understanding and controlling the melting process, it is crucial to elucidate the mechanism of flow and heat transfer in the melting process of frozen droplets.

In this Letter, we report an unconventional melting mode of frozen sessile droplets with the unmelted ice layer deposited at the bottom of the droplet, which seems contrary to our general understanding that the ice should float above the water due to the difference in density. Our study shows that the melting mode is determined by the flow in the melting droplet, specifically, the thermal Marangoni convection. The mechanism of the deposited melting mode is unveiled by exploring the internal flow characteristics, the interaction between the melted fluid and the unmelted ice layer, and the forces on the unmelted ice layer.

The melting processes of frozen droplets on different substrates are shown in FIG.\ \ref{fig:1} (see Section S1 of the Supplementary Material for the experimental details). Once the frozen droplets are heated on different substrates, all melting processes start from the bottom of the droplets. However, the droplet on the single-scale nano-structured (SN) superhydrophobic substrate shows a completely different melting morphology. On the copper (Cu) substrate (in FIG.\ \ref{fig:1}a, Multimedia view) and the hierarchical-scale micro-nano-structured (HMN) superhydrophobic substrate (in FIG.\ \ref{fig:1}b, Multimedia view), the unmelted ice layer floats at the top of the droplet as expected because ice has a lower density than liquid water. In contrast, on the SN substrate, the unmelted ice layer is deposited at the bottom of the droplet, which is opposite to our intuition. In addition to the morphology of the unmelted ice layer, the melting time of the droplet on the SN substrate (in FIG.\ \ref{fig:1}c, Multimedia view) is much less than that on the HMN substrate (in FIG.\ \ref{fig:1}b, Multimedia view), even though these two substrates have the same droplet volume and similar contact angles.

As for the melting process on the SN substrate in FIG.\ \ref{fig:1}c (Multimedia view), from $t = 15$ s to $21$ s, the curvature at the top of the unmelted ice layer remains almost unchanged, whereas the lower interface of the unmelted ice layer retreats continuously. Therefore, the observed increase in the liquid volume above the unmelted ice layer suggests that melting is dominated by the bottom interface of the unmelted ice layer rather than by appreciable melting at the tip.

\begin{figure}
  \centering
  \includegraphics[width=0.96\columnwidth]{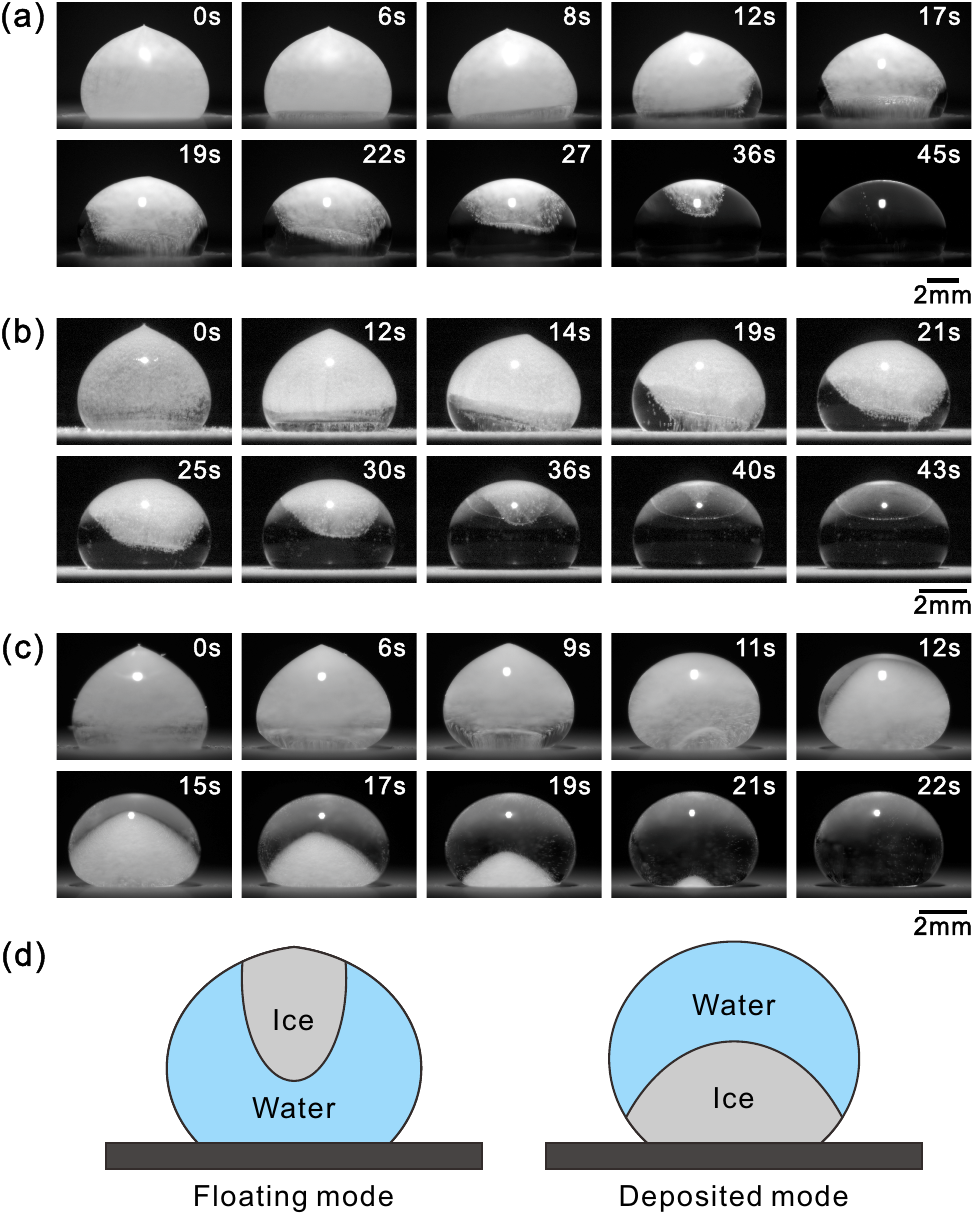}
  \caption{Melting processes of droplets on different substrates: (a) Cu substrate; (b) HMN substrate; (c) SN substrate. Experimental parameters include a heating temperature of 30 $^\circ$C and a droplet volume of 55.6 $\mu$l in panels (b) and (c). (Multimedia available online) (d) Melting modes of droplets on different substrates: floating mode with ice at the top (e.g., on the Cu and HMN substrates); deposited mode with ice at the bottom (e.g., on the SN substrate).}\label{fig:1}
\end{figure}
\begin{figure*}
  \centering
  \includegraphics[width=1.6\columnwidth]{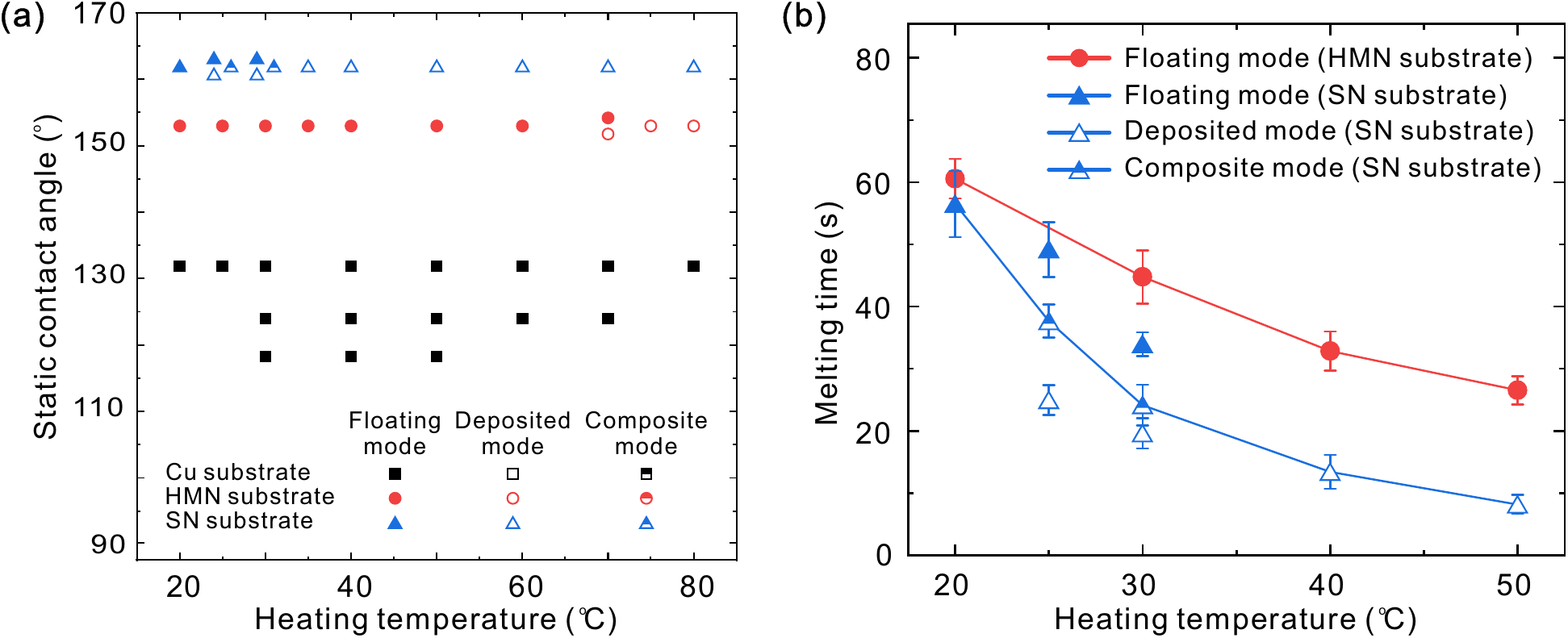}
  \caption{Melting modes (a) and melting times (b) of frozen droplets on different substrates and under different heating temperatures. In panel (a), the maximum droplet volume used is 55.6 $\mu$l on the Cu substrate (which corresponds to the maximum achievable apparent contact angle of 132$^\circ$), and the droplet volume used is 37.1 $\mu$l on the HMN and SN substrates. In panel (b), the blue triangles show the melting times on the SN substrate, among which the solid triangles above the polyline represent the melting times of droplets in the floating mode, and the open triangles below the polyline represent the melting times of droplets in the deposited mode. The points on the polyline (half-filled triangles at the heating temperatures of 25 and 30 $^\circ$C) represent the melting times of droplets in a composite mode (with a transition from the deposited mode to the floating mode, described in Section S3 of the Supplementary Material).}\label{fig:2}
\end{figure*}

To explore the conditions for the occurrence of different melting modes, the melting processes under different heating temperatures are presented in FIG.\ \ref{fig:2}a. For the frozen droplet on the Cu substrate, the melting modes are the same with the heating temperature in the range of 20--80 $^\circ$C. The melting process on the HMN substrate changes with the increase in the heating temperature. At temperatures of 70 and 80 $^\circ$C, the unmelted ice layer deposits at the bottom of the melting droplet, and the frozen droplet has a different melting process compared with that at temperatures from 20 to 60 $^\circ$C. These results suggest that the melting mode is affected by the heating temperature. The same trend can also be found during the melting process on the SN substrate. With the increase in the heating temperature, the unmelted ice layer inside the melting droplet changes from the top to the bottom. However, the difference between the SN and the HMN substrate is that the deposited mode on the SN substrate can be realized at a relatively low temperature of 25 $^\circ$C compared with that of 70 $^\circ$C on the HMN substrate.

In an intermediate range of heating temperatures (25-30 $^\circ$C on the SN substrate), the melting process exhibits a composite mode. In this case, the unmelted ice layer is first deposited at the bottom of the droplet and melts in contact with a thin lubrication film. As the melting proceeds and the internal flow reorganizes, the ice layer eventually detaches from the bottom and rises to the top of the droplet, after which the remaining melting occurs in the conventional floating mode. This deposited-to-floating transition is illustrated and discussed in detail in Section S3 and FIG.\ S6 of the Supplementary Material.

The difference in the melting mode directly affects the rate of heat and mass transfer in the droplet melting process. The melting time is a simple and intuitive parameter to measure the impact of different melting modes, and the melting time under different heating temperatures is plotted in FIG.\ \ref{fig:2}b. On the HMN and SN substrates, the droplet volume is the same, and the contact angle is similar. The experimental result shows that there is a significant difference in the melting time of droplets on these two substrates. 
For example, at a substrate heating temperature of 30 $^\circ$C and for a fixed droplet volume of 37.1 $\mu$L, the average melting time is 19.61 $\pm$ 2.45 s on the SN substrate (deposited mode), in contrast with 44.78 $\pm$ 4.26 s on the HMN substrate (floating mode), corresponding to a reduction in the melting time of about 56\%.
In addition, it is worth noting that the melting time varies with different melting modes at the same heating temperature of 25 or 30 $^\circ$C on the SN substrate. For example, the average melting time at the heating temperature of 25 $^\circ$C is 24.95 $\pm$ 2.61 s in the deposited mode, which is much shorter than 49.19 $\pm$ 4.42 s in the floating mode. Such features of the experimental results indicate the complexity of the melting process.

\begin{figure}
  \centering
  \includegraphics[width=\columnwidth]{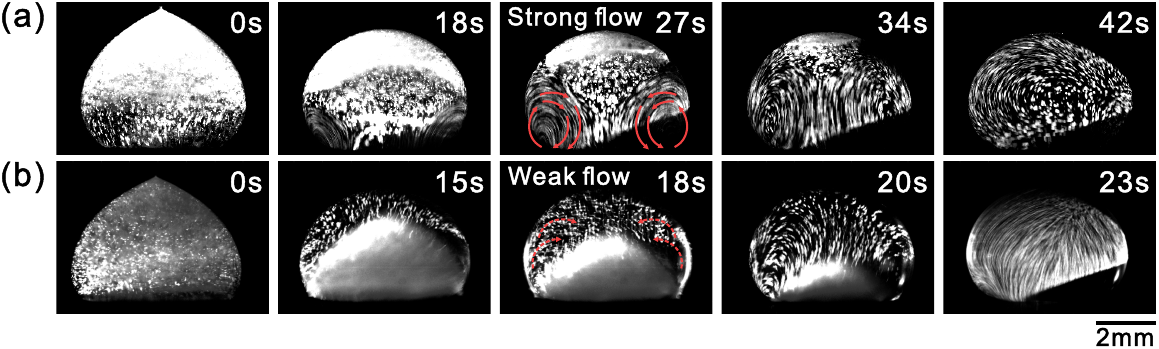}
  \caption{Flow during the melting of frozen droplets on (a) the HMN substrate and (b) the SN substrate. The experiment is conducted with a fixed heating temperature of 30 $^\circ$C and a controlled droplet volume of 37.1 $\mu$l. By overlaying 40 consecutive images acquired at 100 fps, the flow field information inside the droplet can be intuitively obtained.
}\label{fig:S3}
\end{figure}

To understand the mechanism for different melting modes of the unmelted layer, the internal flow of the melted liquid in the droplet and its effect on the morphology change of the unmelted ice layer are experimentally explored. The temperature difference in the droplet induces flow during the melting process, as shown in FIG.\ \ref{fig:S3}.  On the HMN substrate, there are two regular flow vortices within the melting droplet (see FIG.\ \ref{fig:S3}a). The melting flow intensity is relatively strong, especially near the gas-liquid interface. Due to the difference in the flow speed at different regions inside the droplet, the melting speed at the bottom of the unmelted ice layer is relatively slower, while the melting speed on the left/right sides of the unmelted ice layer is much faster. In contrast, for the melting process on the SN substrate, the flow intensity in the melted area is much weaker than that on the HMN substrate, and regular flow can only be seen near the left/right sides of the unmelted ice layer (see FIG.\ \ref{fig:S3}b). The difference in the flow field seems to contradict the difference in the melting time: in the deposited melting mode, which has a very weak flow, the frozen droplet actually has a very quick melting process. This means that there is a different mechanism that leads to more efficient heat transfer between the ice layer and the heating wall, and not because of the convection.

\begin{figure}
  \centering
  \includegraphics[width=0.9\columnwidth]{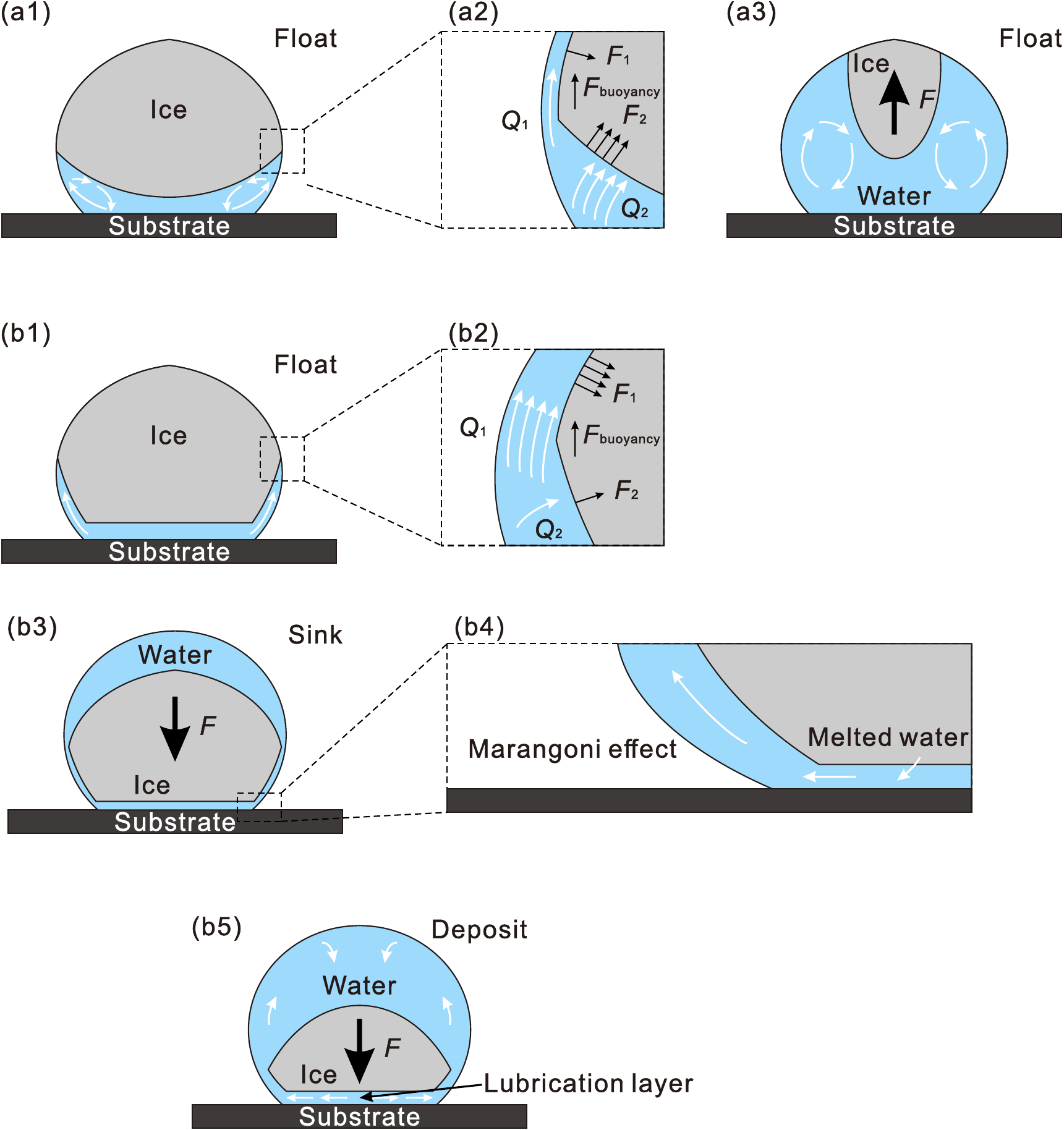}
  \caption{Schematic diagrams of the melting modes and force analysis. (a) Schematics diagram of the floating mode, including the early stage (a1), the force analysis (a2), and the stable stage (a3). (b) Schematics diagram of the deposited mode, including the early stage (b1), the force analysis (b2), the liquid film (b3), the Marangoni effect (b4), and the stable stage (b5).  
}\label{fig:3}
\end{figure}

To further explore the mechanisms for different melting modes, the melting process under different contact angles, different heating temperatures, and different particle concentrations is studied (see Section S2.2 of the Supplementary Material). The contact angle directly affects the morphological characteristics of the unmelted ice layer, while the heating temperature and the particle concentration indirectly affect the morphological characteristics of the unmelted ice layer by changing the melting flow intensity. According to these experimental results, schematic diagrams of the droplet melting are shown in FIG.\ \ref{fig:3}. In the early stage of droplet melting, only a small melted area appears at the bottom of the droplet in both melting modes (which can be seen at $t = 14$ s in FIG.\ \ref{fig:1}b and $t = 9$ s in FIG.\ \ref{fig:1}c). Hence, Marangoni convection gradually forms \cite{Chu2019reicing, Lei2015melting} and has a high flow intensity in this stage. Due to the differences in the robustness of micro/nano-structures, different amounts of particles detach from the substrate surface, having different degrees of impact on the intensity of Marangoni convection (see Section S2.2 and FIG.\ S5 in Supplementary Material). In addition, different heating temperatures can also change the intensity of Marangoni convection (see Section 2.2 and FIG.\ S4 in Supplementary Material). These two aspects ultimately lead to the difference in the shape of the unmelted ice layer (see FIG.\ \ref{fig:1}). The longer dimension along the gas-liquid interface direction in FIG.\ \ref{fig:3}(b1) can lead to higher flow intensity than that in FIG.\ \ref{fig:3}(a1). Therefore, the forces of these two melting modes in the three-phase area (i.e., the contact area of air, water, and ice) are different. For the melting process in the two modes, schematic diagrams of the three-phase area are shown in FIGs.\ \ref{fig:3}(a2) and \ref{fig:3}(b2). Importantly, a thin liquid film forms between the air and the upper ice surface due to thermomolecular pressure generated during surface melting, which induces interfacial mass transport under lateral temperature variations \cite{10}. 
Here, the flow due to the Marangoni effect has two characteristic paths: one is to bypass the unmelted ice layer and flow to the area above the ice layer, and the other is to flow to the area below the unmelted ice layer. These two flow paths further lead to two different motions of the unmelted ice layer: the flow to the area above the unmelted ice layer tends to press the ice layer downward, while the flow to the area below the unmelted ice layer tends to push the ice layer upward.

For the melting process with a relatively low flow intensity, more melted fluid would flow toward the bottom of the ice layer (see FIG.\ \ref{fig:3}(a2)), that is, $Q_2 > Q_1$, where $Q_1$ and $Q_2$ represent the flow rates of melted fluid toward the top and bottom of the ice layer, respectively. In the vertical direction, the force component of $F_2$ is greater than that of $F_1$, where $F_1$ and $F_2$ are the forces exerted on the ice layer by the aforementioned flows $Q_1$ and $Q_2$. Therefore, the combined force of $F_2$ and $F_\text{buoyancy}$ in the vertical direction must be greater than $F_1$ in the vertical direction. As a consequence, the ice layer moves to the top of the droplet. For the melting process with relatively high flow intensity, more melted fluid would flow bypass the unmelted ice layer and toward the top of the ice layer, that is, $Q_1 > Q_2$, as illustrated in FIG.\ \ref{fig:3}(b2). The force due to the flow effect satisfies that the component force by $F_1$ is greater than that of $F_2$. Therefore, the downward effect provided by the combined force of $F_2$ and $F_\text{buoyancy}$ may be less than the upward effect provided by $F_1$, and the unmelted ice layer will be forced to deposit at the bottom under the effect of melted fluid (as shown in FIG.\ \ref{fig:3}(b3)).

The aforementioned analysis explains the underlying mechanism by which the unmelted ice layer can sink. However, in order to achieve the deposited mode, it is necessary to keep the unmelted ice layer stable at the bottom of the droplet because of the change in the melting flow (see Section S2.1 of the Supplementary Material). As the melting process continues, a very thin liquid film appears between the unmelted ice layer and the heated wall surface. Although the presence of the film is difficult to observe directly from FIG.\ \ref{fig:1}c, it can be inferred based on the thermal characteristics at the contact area. Because the main melting area is the bottom of the unmelted ice layer and the liquid produced after the melting is transferred to the top of the droplet by the Marangoni effect, there is continuous flow within this liquid film [as shown in FIG.\ \ref{fig:3}(b4)]. According to the experimental images, the thickness of the liquid film $h$ is far less than the diameter of the liquid film $d$. The Reynolds number $Re$ of the melting flow inside the liquid film can be calculated by $Re = 2 h \bar{u}/v$, where $\bar{u}$ is the average flow velocity, and $\nu$ is the kinematic viscosity of water. Due to the fact that the melting area is mainly distributed at the bottom of the unmelted ice layer, almost all melted fluids flow out from the liquid film, so $\bar{u}$ can be calculated by $\bar{u} =V/(\pi d h t)$, where $V$ is the droplet volume, and $t$ is the melting time. The product of the dimensionless liquid film thickness $h/d$ and the Reynolds number $Re$ is far less than 1, satisfying the requirements of lubrication approximation. That is, the inertia effect in the liquid film can be ignored, and the flow is mainly dominated by viscous forces. Thus, significant flow resistance is generated during the separation of these two solid interfaces, which may suppress the upward movement of the unmelted ice layer (as shown in FIG.\ \ref{fig:3}(b5)).

To complement the mode map in FIG.\ \ref{fig:2} and to provide a concise theoretical basis for FIG.\ \ref{fig:3}, we introduce a simplified scaling model that captures the competition among thermocapillary driving, buoyancy of the ice layer, and viscous lubrication at the ice-substrate interface. We define a thermocapillary-to-buoyancy parameter
\begin{equation}
  \Pi = \frac{\left|\mathrm{d}\sigma/\mathrm{d}T\right|\Delta T}{\Delta\rho g L^{2}}
\end{equation}
where $\Delta T$ is the substrate-to-melting-front temperature difference and $L$ is a characteristic droplet length scale. Large $\Pi$ favors a stronger interfacial bypass flow feeding melted liquid above the ice layer $Q_1>Q_2$, consistent with the deposited mode in FIG.\ \ref{fig:3}(b). The stability of the deposited ice is further supported by viscous lubrication in the thin film beneath the ice. Using $\bar{u}=V/(\pi d h t)$, the ratio of viscous to buoyancy pressure scales can be estimated as
\begin{equation}
  \Lambda \sim \frac{\mu \bar{u} L}{h^{2}\Delta\rho g H}
\end{equation}
The deposited mode is expected when $\Pi$ and $\Lambda$ are sufficiently large, which rationalizes the experimental trends in FIG.\ \ref{fig:2} with increasing heating temperature, large contact angle, and reduced particle-induced damping. Increasing the heating temperature increases $\Delta T$, thus $\Pi$, promoting $Q_1>Q_2$. A larger apparent contact angle increases droplet aspect ratio and narrows the interfacial transport corridor, effectively lowering the threshold of $\Pi$ required for deposition. Reduced particle detachment weakens interfacial damping and strengthens thermocapillary flow, again increasing the effective $\Pi$, which explains why the SN substrate can realize the deposited mode at lower temperatures than the HMN substrate.

The mode map and melting-time measurements (FIG.\ \ref{fig:2}), and the parametric studies of contact angle, heating temperature, and particle concentration (FIGs.\ S4 and S5), together with lubrication-based estimates of Reynolds number and viscous and buoyancy forces in the thin film, quantitatively support the flow configurations illustrated in FIG.\ \ref{fig:3}. Dimensionless estimations show ($Bo < 1$, $Ma \sim 10^4-10^5$, $Ra \sim 10^3-10^4$, and $(h/d)Re \ll 1$), indicating a capillary-dominated droplet with Marangoni-dominated interfacial flow and lubrication-dominated film dynamics. A full three-dimensional numerical simulation that resolves the free surface, the moving ice-water interface, and the lubrication film is an important direction for future work.

A high flow intensity and narrow flow area at the gas-liquid interface of a melting droplet are important for the deposition of the unmelted ice layer at the bottom. The high flow intensity can be achieved by increasing the heating temperature of the substrate and reducing the particle concentration of the droplet. The narrow flow area at the gas-liquid interface requires an increase in the flow intensity, while ensuring that the droplet has a larger contact angle to increase its aspect ratio. More details on the conditions for the deposition of the unmelted ice layer during the melting process are shown in Section S4 of the Supplementary Material.

In summary, an unconventional melting mode with the unmelted ice layer deposited at the bottom of the droplet is identified.
SN substrates exhibit fundamentally different melting behavior, characterized by uniform melting on the bottom and minimal fluid flow. The unmelted ice layer deposits at the bottom of the melting droplet, even though the density of ice is lower than that of water. As the heating temperature rises and the particle concentration decreases, the melting flow along the gas-liquid interface becomes stronger. A higher contact angle is beneficial for the formation of relatively narrow flow areas. All of these can guide the melted fluid flow to the top of the unmelted ice layer and then make the ice layer move downward. According to the lubrication theory, the existence of a very thin liquid film makes it difficult to separate the two solid interfaces due to the dominance of viscous forces in the melting flow. Therefore, the unmelted ice layer can stably deposit at the bottom of the droplet, i.e., the deposited mode. This melting mode is helpful to further unleash the anti-icing and deicing capabilities of superhydrophobic surfaces.

See Supplementary Material for the experimental details, characteristics of the melting process, transition of the melting mode, and conditions for the deposited mode.

This work was supported by the National Natural Science Foundation of China (Grant No. 52176083).

\bibliography{MeltingBottom}

\end{document}